%% file: main.tex
\documentclass[conference]{IEEEtran}
\IEEEoverridecommandlockouts
\usepackage{cite}
\usepackage{amsmath,amssymb,amsfonts}
\usepackage{algorithmic}
\usepackage{graphicx}
\usepackage{textcomp}
\usepackage{xcolor}
\usepackage{hyperref}

\def\BibTeX{{\rm B\kern-.05em{\sc i\kern-.025em b}\kern-.08em
    T\kern-.1667em\lower.7ex\hbox{E}\kern-.125emX}}
\begin{document}

\title{Gem5Pred: Predictive Approaches For Gem5 Simulation Time}

\author{\IEEEauthorblockN{Tian Yan}
\IEEEauthorblockA{
\textit{University of Notre Dame}\\
Notre Dame, USA \\
tyan@nd.edu}
\and
\IEEEauthorblockN{Xueyang Li}
\IEEEauthorblockA{
\textit{University of Notre Dame}\\
Notre Dame, USA \\
xli34@nd.edu}
\and
\IEEEauthorblockN{Sifat Ut Taki}
\IEEEauthorblockA{
\textit{University of Notre Dame}\\
Notre Dame, USA \\
staki@nd.edu}
\and
\IEEEauthorblockN{Saeid Mehrdad}
\IEEEauthorblockA{
\textit{University of Notre Dame}\\
Notre Dame, USA \\
smehrdad@nd.edu}
}

\maketitle

\begin{abstract}
Gem5, an open-source, flexible, and cost-effective simulator, is widely recognized and utilized in both academic and industry fields for hardware simulation. However, the typically time-consuming nature of simulating programs on Gem5 underscores the need for a predictive model that can estimate simulation time. As of now, no such dataset or model exists. In response to this gap, this paper makes a novel contribution by introducing a unique dataset specifically created for this purpose. We also conducted analysis of the effects of different instruction types on the simulation time in Gem5. After this, we employ three distinct models leveraging CodeBERT to execute the prediction task based on the developed dataset. Our superior regression model achieves a Mean Absolute Error (MAE) of 0.546, while our top-performing classification model records an Accuracy of 0.696. Our models establish a foundation for future investigations on this topic, serving as benchmarks against which subsequent models can be compared. We hope that our contribution can simulate further research in this field. The dataset we used is available at \url{https://github.com/XueyangLiOSU/Gem5Pred}.

\end{abstract}

\input{sections/intro}
\input{sections/Background}

\input{sections/method}
\input{sections/Experiments}
\input{sections/Results_and_Discussion}

\input{sections/Conclusion}

\bibliographystyle{unsrt}
\bibliography{sections/ref}
\end{document}

%% file: sections/intro.tex
\section{Introduction}

Gem5~\cite{10.1145/2024716.2024718} is a highly flexible, full-system simulator that harmoniously merges the strengths of two distinct projects: M5, a full-system simulator \cite{dreslinski2008m5}, and GEMS, a focus on memory systems \cite{martin2005multifacet}. It facilitates the exploration of various architectural features of multiprocessors by offering a broad range of CPU models, system execution modes, and memory system models \cite{binkert2011gem5}. Moreover, Gem5 supports a plethora of commercial ISAs, including ALPHA, ARM, x86, SPARC, PowerPC, and MIPS, enhancing its modularity and ease of CPU model interchange \cite{butko2012accuracy}. Being open-source, Gem5 is accessible to a diverse group of users and allows for unrestricted modification, usage, and distribution of the software. These attributes have contributed to its widespread adoption in both academia and industry for computer architecture research. However, the very traits that make Gem5 advantageous - its high modularity, customization, and provision of an accurate, detailed, and flexible simulation environment - can also lead to prolonged simulation times due to its sequential nature \cite{qureshi2019gem5}. As such, executing complex programs on Gem5 can be quite time-consuming, underscoring the need for a model capable of predicting Gem5's simulation time prior to program execution.

To our knowledge, there is no existing literature that presents or discusses a model dedicated to predicting the simulation time of Gem5. As such, several challenges must be overcome in order to develop such a predictive model. Firstly, there is currently no available dataset to train this model, necessitating the creation of a new dataset. Secondly, the absence of a standard baseline complicates the assessment of the model's performance and comparison with other potential models. Thirdly, there are no readily available models specifically designed for similar prediction tasks. Another significant challenge is the broad range of Gem5 simulation results, spanning from microseconds (1e-6 seconds) to seconds. This wide spectrum could add complexity to the task of devising a predictive model.

In this paper, we first investigate how different types of instructions will affect the simulation results (the ``SimSeconds'') of Gem5. This is accomplished by generating diverse program types and simulating them on Gem5. Subsequently, we address the challenges previously mentioned and introduce three distinctive predictive models for the prediction of Gem5 SimSeconds, leveraging the capabilities of large language models.


The structure of this paper is as follows: Section II provides the background necessary for understanding our models. Section III delves into the specifics of our dataset and the methodology employed. Experimental procedures are detailed in Section IV, while Section V discusses the outcomes of these experiments. Lastly, Section VI encapsulates the conclusions drawn from this study.

%% file: sections/Background.tex
\section{Background}

\subsection{Large Language Models}
In recent years, advancements in Natural Language Processing (NLP) tasks have been markedly propelled by the advent of pre-trained large language models such as ELMo~\cite{peters2018dissecting}, GPT~\cite{radford2018improving}, BERT~\cite{devlin2018bert}, and RoBERTa~\cite{liu2019roberta}. These models have the capability of doing both embedding and classification. The impressive performance of pre-trained models in NLP has catalyzed the development of multi-modal pre-trained models, which are trained on bimodal data, such as language-image pairs, using bimodal self-supervised objectives. One notable example of a bimodal pre-trained model is CodeBERT~\cite{feng2020codebert}, designed to facilitate natural language (NL) and programming language (PL) understanding. CodeBERT captures the semantic relationship between NL and PL, resulting in the creation of general-purpose representations that effectively serve NL-PL understanding and generation tasks. Just like other language models, it can be utilized in various applications, such as doing sequence classification directly from code scripts or converting code scripts into embedded matrices, as demonstrated later in our work.

\subsection{Support Vector Regression}

Regression analysis is a valuable tool for examining the relationship between a dependent variable and one or more predictor variables. In the context of learning a regression function that maps input predictor variables to observed response values, Support Vector Regression (SVR) proves to be beneficial. It strikes a balance between model complexity and prediction error~\cite{smola2004tutorial,zhang2020support}, making it particularly suitable for high-dimensional data analysis. While SVR is an extension of the Support Vector Machine (SVM) classification algorithm \cite{boser1992training}, it specifically addresses regression problems, enabling the estimation of continuous-valued functions rather than binary outputs. One of the advantages of SVR is its ability to handle nonlinear regression problems by employing a kernel \cite{ben2008support}. Unlike other regression models that often require assumptions about the validity of a specific model, SVR trains a model to capture the underlying significance of variables in describing the relationship between input and output \cite{zhang2020support}. This feature sets SVR apart from alternative regression methods.

In the following paragraph, we will outline how we integrated large language models, particularly CodeBERT, with Support Vector Regression (SVR) to address the challenges at hand and enable the prediction of simulation time on Gem5 based on our generated dataset using the original code script as input.

%% file: sections/method.tex
\section{Dataset and Methodology}


Gem5 simulation time can depend on many factors. To cover a wide variety of use cases, we initially generated 150 C language programs with basic arithmetic and memory operations. With the help of ChatGPT~\cite{brown2020language}, we generated source codes for the samples for our initial testing. Next, we tested the initial C programs by simulating them on Gem5. It generates a detailed statistics file for each program simulated~\cite{binkert2011gem5}. We carefully analyzed these statistics files to understand the behavior of each program and its impact on simulation time. Based on the insights gained from the simulation statistics, we made further adjustments to each program to ensure that the generated statistics served their intended purpose. During the modification process, we experimented with different input sizes, such as varying the number of memory operations, dependent variables, nested loops, iterations per loop, and function calls that introduced branch operations. These variations allowed us to observe the specific effects of each factor on the Gem5 simulation. Ultimately, we expanded our sample size to encompass 280 programs with the configuration described in Table~\ref{tab:gem5_config}, thereby increasing the diversity and coverage of our dataset.

\begin{table}[]
\centering
\caption{Gem5 simulator configuration.}
\label{tab:gem5_config}
\begin{tabular}{c|c}
\hline
\textbf{Parameter}                 & \textbf{Configuration}   \\ \hline
Architecture              & x86             \\ 
CPU type                  & TimingSimpleCPU \\ 
L1 data cache size        & 64 kB           \\ 
L1 instruction cache size & 16 kB           \\ \hline
\end{tabular}
\end{table}

\subsection{Simulation Dataset}
Below, we provide a comprehensive explanation of the characteristics and details of each type of program in our dataset.

\noindent \textbf{Basic arithmetic operations:} To see the impact of very simple arithmetic operations--such as additions, subtractions, multiplications, and divisions--we tested a few very small programs with such operations. The impact on simulation time was minimal as expected. Each program simulated within about 0.0003 seconds.

\noindent \textbf{Basic memory access operations:} The basic memory access operation programs contained arrays of different datatypes. Each program had instructions for writing to the memory first and reading some values from the stored array. For these programs, the simulation time varied depending on the number of memory operations incurring within each corresponding program. These programs help us understand the impact of memory operations on Gem5 simulation.

\noindent \textbf{Matrix multiplications:} We generated and tested programs with sizes of 2-dimensional and 3-dimensional matrices. Each program with different matrix operations such as addition and multiplications. These programs helped us to evaluate the impact of nested loops on Gem5 simulation time. These programs had time complexities of $O(n^2)$ and $O(n^3)$ depending on the dimensions. Memory access operations also varied for each program to see the combined effect. Depending on the sizes and dimensions of the matrices--which resulted in varying numbers of nested loops and the number of iterations per loop.

\noindent \textbf{Sorting algorithms:} We tested the following sorting algorithms: bubble sort, insertion sort, selection sort, merge sort, and quick sort. These algorithms helped us understand the impact of general use cases. The time complexity of the sorting algorithms ranged from $O(n^2)$ to $O(nlogn)$. We also experimented with different input sizes for each algorithm. The simulation time for these programs depended on the time complexity and input size.

\noindent \textbf{Searching algorithms:} We have also tested two searching algorithms: binary search and Fibonacci search. Both have a time complexity of $O(logn)$. As such, these programs were faster to simulate compared to the sorting algorithms. We experimented with different input sizes as well to see the impact on simulation time.

\noindent \textbf{Hashing algorithms:} We have also tested some hashing algorithms as well. For the hashing applications, we have tested the Blake and the adler32 hashing algorithms with different input sizes.

\begin{table}[]
\caption{Simulation dataset distribution}
\label{tab:sim_dist}
\begin{tabular}{|l|l|l|l|l|l|l|l|}
\hline
\textbf{Metric}          & \textbf{Min}    & \textbf{25\%}    & \textbf{50\%}    & \textbf{75\%}    & \textbf{Max}    & \textbf{Mean}   & \textbf{Std}    \\ \hline
Length     & 85     & 391    & 831    & 1183   & 2740   & 897.5 & 564.9 \\ \hline
Time(s) & .0003 & .0003 & .0009 & .1051 & 6.3892 & .2398 & .7570 \\ \hline
\end{tabular}
\end{table}

After generating C programs for different applications and simulating them using Gem5, we started finalizing our dataset for the model predictions. The final dataset consists of 280 source codes of different C programs and their corresponding simulation time. Table~\ref{tab:sim_dist} presents the distribution of the code lengths and simulation times of our finalized dataset. For the classificaion model, we pre-classified the programs based on the simulation time, which will be explained later. By analyzing the dataset, we realized that the simulation time for a program depends on the following factors:
\begin{itemize}
    \item The number of memory access operations significantly increases the simulation time.
    \item The number of iterations within each loop contributes to higher simulation time.
    \item The matrix dimensions exponentially increase the simulation time.
    \item We have seen an impact of the data dependency on simulation time; however, the impact is not as significant as the aforementioned factors.
\end{itemize}

\subsection{Methodology}
As previously discussed, one of the prominent challenges we confronted was the excessively wide range and imbalancement of the generated data. Our program produces simulation times that vary drastically, ranging from microseconds (1e-6) to full seconds with more data concentrated in a short time, posing significant hurdles to accurate prediction. To address this, we devise two strategies. The first involves applying a base-10 logarithmic transformation to the simulation times, thereby increasing the data density. Our second approach relies on a preliminary classification of the dataset based on their magnitude. For simulation times greater than one, we treated each second as a discrete magnitude. Conversely, for values less than one, we considered each digit following the decimal point as a distinct magnitude. The first strategy lends itself to predicting continuous values, while the second strategy proves effective for forecasting a specific range of values. 

One model we widely used in our methods is CodeBERT. CodeBERT has the same model structure model as RoBERTa\cite{liu2019roberta} base model, which is an improved version of BERT model\cite{devlin2018bert}. CodeBERT is different from RoBERTa in several ways. First, they sourced their training data from Github, and paired the function and its documentation as a bimodal training data point, and used the function alone as a unimodal training data point. Hence, they have both natural language and codes in training data, and they include a [SEP] token between function and documentation for the bimodal data type. Further, they used bimodal data to train the masked language modeling task and unimodal data to train replaced token detection task, which allow them to fully utilize both data types.

Our models are derived from CodeBERT~\cite{feng2020codebert} in three distinct ways: 1) Utilizing CodeBERT~\cite{feng2020codebert} for embedding, followed by Support Vector Regression (SVR) applied to the embedded space for continuous value prediction. 2) Fine-tuning CodeBERT to serve as a predictive model for direct continuous value prediction. 3) Fine-tuning CodeBERT to function as a predictive model for simulation time range classification. Aside from these models, we also consider two very simple models as the baseline. In the following paragraphs, we delve into the specifics of these approaches.

\subsubsection{Baseline}
\noindent \textbf{Code length as a predictor: }Considering code length can be an effective predictor for simulation time and lengthy codes tend to run longer, we used code length as the only predictor to predict the code simulation time and use its performance as the first baseline. Specifically, for each code sample, we count its number of characters and use it as the only predictor for simulation time and train an SVR model to do prediction.

\noindent \textbf{Always predict median: }Considering we have a very skewed time distribution in the dataset, it's likely that a model that always predicts the median simulation time, or always predicts the most popular time class can yield seemingly good performance. Thus, we use the median simulation time as the predicted time for all testing samples and use it as a baseline. And models that can truly learn the pattern of data should outperform this baseline easily.

\subsubsection{CodeBERT + SVR}
In this model, we first process each sample with a tokenizer, then we feed the tokenized data into the pre-trained CodeBERT model. After this, we take the last layer pooled output of [CLS] token as the embedding for this sample, which gives us a 768$\times$ 1 feature vector. We treat each element in the embedding vector as a predictor for the next-step SVR model, resulting in a total of 768 predictors. As for the response variable, we used the logarithm (base 10) of the simulation time. Finally, we train an SVR model based on the predictors and the response variable. The purpose of this SVR model was to predict the logarithm of the simulation time based on the provided predictors.

\subsubsection{CodeBERT (Regression)}
In this model, we directly finetune the CodeBERT model to do prediction instead of using another model. Specifically, we add another randomly initialized fully connected layer on top of the last layer [CLS] pooled output layer, which maps the 768$\times$ 1 output to 1 single response variable prediction. Due to the small amount of data, we only updated the [CLS] pooling layer and the newly added fully connected layer, and hold all the layer parameters unchanged. The response variable is still the logarithm (base 10) of simulation time.

\subsubsection{CodeBERT (Classification)}
This model's fundamental process mirrors that of the previous model, which employed CodeBERT for regression, with one key difference. Instead of applying a base-ten logarithmic transformation, we segregated the resulting simulation time into nine distinct categories: $(0, 10^{-3})$, $[10^{-3}, 10^{-2})$, $[10^{-2}, 10^{-1})$, $[10^{-1}, 1)$, $[1, 2)$, $[2, 3)$, $[3, 4)$, $[4, 5)$, $[5, + \infty)$. They are labeled as class 0, 1, 2, 3, 4, 5, 6, 7, and 8, respectively.  This approach essentially turns the task into a multi-label classification problem, with CodeBERT serving as the core predictive tool. The pre-classification of labels serves to enhance data density, thereby facilitating a more streamlined prediction process for our model.

In summary, we implemented two distinct workflows for the model. The first involves utilizing CodeBERT\cite{feng2020codebert} as an embedding model and applying machine learning models to the embedded space for predictive purposes. The second approach leverages CodeBERT directly for prediction. Within this latter workflow, we experimented with both the regression model and the classification model.

%% file: sections/Experiments.tex
\section{Experiments}
As our dataset is limited, we use 80\% data to train our model and use the remained 20\% for the test, which result in 224 training samples and 56 testing samples. 
Below are the detailed experiment settings for the three models.

\subsubsection{CodeBERT + SVR}
In the CodeBERT part, we directly used the pre-trained parameters from CodeBERT to get all embeddings. And for SVR model, we fit the model with SVR function from Python sklearn package and choose an $l_2$ regularization parameter of 1 and $\epsilon=0.2$ in the epsilon-SVR model. All other parameters follow the default setting of sklearn. 

\subsubsection{CodeBERT (Regression)}
For the regression CodeBERT model, we fine-tuned the last pooling layer of [CLS] token of the original model, together with a linear projection layer that we add on top of the last pooling layer. We used AdamW optimizer with $5e-5$ learning rate. The batch size is set to 16, and the model is finetuned for 30 epochs on a machine with a single Intel L5520  @2.27GHz Xeon CPU and 72.0 GB memory.           

\subsubsection{CodeBERT (Classification)}
For the Classification CodeBERT model, we fine-tuned the last layer of the original CodeBERT model, just like what we did in Regression CodeBERT. We used the Adam optimizer~\cite{kingma2014adam} with $1e-5$ learning rate. The batch size is set to 16 and the warmup step is set to 5. This model is fine-tuned for 20 epochs on a machine with 24 Intel E5-2680 v3 @ 2.50GHz Xeon CPUs and took about 12 hours. 

%% file: sections/Results_and_Discussion.tex
\begin{table}
\centering
\caption{MAE/Accuracy of different models on our dataset}\label{result}
\begin{tabular}{c|c}
\hline
{\bfseries  Model}&  {\bfseries  MAE } \\
\hline
Baseline (code length) &  1.025  \\
Baseline (median value) & 0.996 \\
CodeBERT embedding + SVR & {\bfseries 0.546} \\
CodeBERT-based Regression & 0.853 \\
\hline
\hline
{\bfseries  Model}&  {\bfseries  Accuracy }\\
\hline
CodeBERT-based Classification & 0.694\\
\hline
\end{tabular}
\end{table}

\section{Results and Discussion}
For all regression problems, we employ the mean absolute error (MAE) as the evaluation metric. The MAE is calculated as the mean of the absolute differences between the true and predicted values. It provides a measure of the average magnitude of the errors in the predictions. On the other hand, for classification problems, we use the overall accuracy as the evaluation metric. Accuracy is determined by calculating the total number of samples that have been correctly classified and dividing it by the total number of samples in the dataset. This metric provides an assessment of the overall correctness of the classification predictions. Their performances are presented in Table~\ref{result}.

\subsection{Baseline}
In the baseline approach where we solely utilize code length as a predictor, we achieved a mean absolute error (MAE) of 1.025. When we employed the baseline approach of predicting the median value for all samples, we achieved a slightly improved MAE of 0.996.

The unexpected outcome reveals that code length alone is not as influential of a predictor as initially anticipated, and it underperforms compared to the median predictor. This could be attributed to the fact that parameters such as loop length, matrix size, and memory access times tend to have a more deterministic impact on simulation time. These factors are not directly related to code length. Code length may play a more significant role in larger programs that require days or weeks to execute, but such programs are not included in our dataset.

\begin{figure}
\includegraphics[width=8cm]{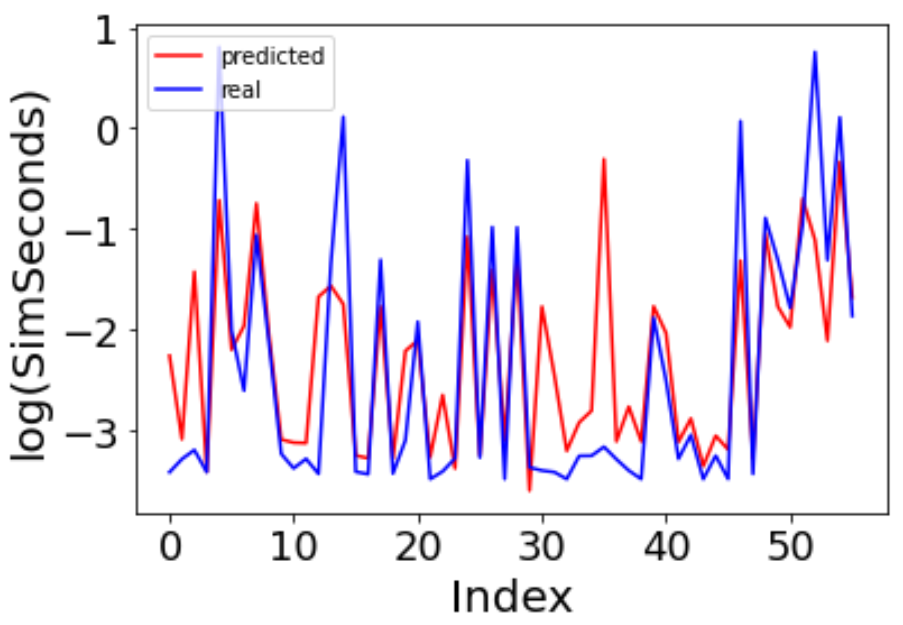}
\caption{Comparison between real labels and predicted labels generated by SVR (regression)} \label{SVR_pred}
\end{figure}

\subsection{CodeBERT + SVR}
In the model where we use CodeBERT embedding and train a SVR model, we could improve the MAE to 0.546 compared with the baseline. A detailed comparison between the true and predicted labels is visually depicted in Figure~\ref{SVR_pred}.

In the figure, the model demonstrates strong performance for small to median true values, accurately predicting the simulation time. However, it tends to generate larger errors for large true values, indicating a tendency to underestimate the time required for simulations when the true time is considerably long. Additionally, there are several samples where the true time is small but have a relatively large predicted time. Upon investigating these specific samples, we observed that they involve explicitly writing out the values of multiple variables. We will delve into possible explanations for this phenomenon in the Error Analysis section.

\begin{figure}
\includegraphics[width=8cm]{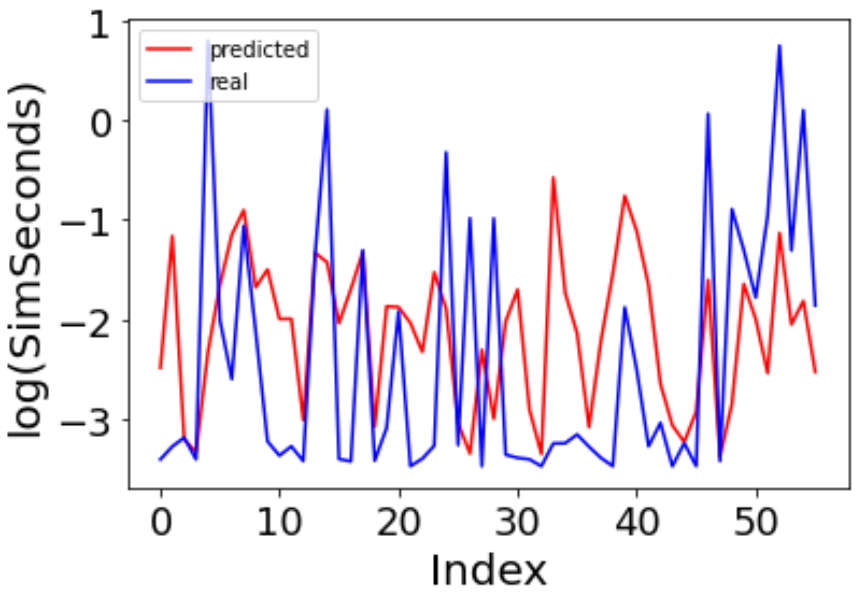}
\caption{Comparison between real labels and predicted labels generated by CodeBERT(regression)} \label{finetune_reg_pred}
\end{figure}

\subsection{CodeBERT (Regression)}
In the model where we finetune the CodeBERT model, we could improve the MAE to 0.853 compared to the baseline. The detailed comparison between the true and predicted labels is visually represented in Fig.~\ref{finetune_reg_pred}. 

Interestingly, we observed that despite having a larger number of parameters and greater complexity, CodeBERT did not outperform the SVR model. This can also be seen in Fig.~\ref{finetune_reg_pred}, where CodeBERT exhibits accurate predictions for certain data points, but overall, it demonstrates a certain degree of bias. This indicates that the model is not effectively learning the underlying data patterns. One possible explanation for this suboptimal performance is the lack of nonlinearity in the last layer of the fine-tuned model. Due to the nature of the regression problem and the absence of activation functions or nonlinearity in the last layer, the model might struggle to capture complex relationships in the data. This limitation may have impacted its predictive performance compared to the SVR model.

\begin{figure}
\includegraphics[width=8cm]{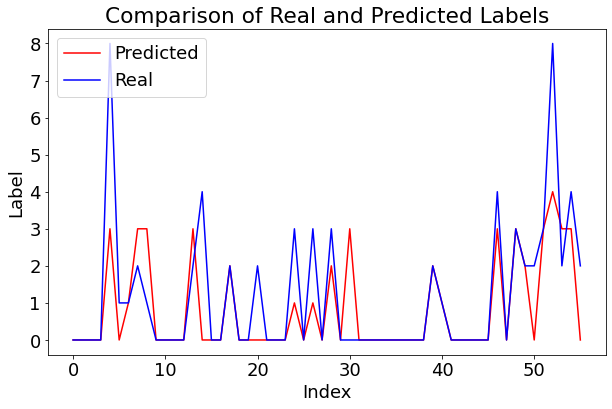}
\caption{Comparison between real labels and predicted labels generated by CodeBERT (classification)} \label{class}
\end{figure}

\subsection{CodeBERT (Classification)}

When we apply the CodeBERT for the Classification task, we obtain an accuracy of \textbf{0.694} on the testing dataset. The detailed comparison between the true and predicted labels is visually represented in Fig.~\ref{class}. 

The illustration clearly demonstrates that similar to other models, this particular model performs well when handling input programs with shorter simulation times. Additionally, upon comparing Figure~\ref{class} with Figure~\ref{finetune_reg_pred} and Figure~\ref{SVR_pred}, it becomes apparent that for shorter simulation times, this model can actually provide more accurate predictions compared to the previous two models. This is evident from the close overlap between the predicted curve and the actual curve in the lower part of the figure. The improved performance of this model for shorter simulation times may be attributed to the fact that, as part of our approach, we pre-classified the simulation times into specific ranges. Consequently, the model only needs to predict a rough range instead of a precise value. As a result, any small prediction differences between the real value and the predicted value, which may have been noticeable in the previous two models, can be effectively ignored. 
Nevertheless, its performance appears to dwindle when predicting inputs with ling simulation time. Despite this, it is noteworthy that this retains the capacity to discern overarching patterns. For instance, even when it mispredicts for inputs with larger simulation time values, the model often estimates a value greater than the typical predictions, suggesting an innate understanding of the underlying pattern. 

\subsection{\textbf{Error Analysis}}
Our models appear to generate two distinct types of errors. The first involves large actual values being predicted as small. This phenomenon, evident across all three models, sees a higher accuracy for smaller simulation time values and a decrease for larger values. This trend can be primarily associated with the inherent imbalance in our dataset, as delineated in Table~\ref{tab:sim_dist}. The dataset predominantly consists of values less than one second, which could complicate the model's ability to accurately predict larger values. Consequently, we hypothesize that the performance of our classification model could be enhanced with access to a larger, more balanced dataset in the future. Another possible explanation for this discrepancy could be that CodeBERT~\cite{feng2020codebert} struggles to accurately capture subtle but crucial alterations in the C language code scripts, such as changes in matrix dimensions. To investigate this further, we examined the two programs with the highest value in Fig.~\ref{class}. Despite their code length being less than 80 lines, one of the programs features high-dimension matrices, while the other includes extensive for-loop dimensions. A minor alteration - reducing either the matrix dimension or the for-loop length to a typical range - significantly improves our model's prediction accuracy for these programs. Hence, we put forth the hypothesis that by translating the C language code into a more foundational language like X86 assembly, the prediction process might be simplified for CodeBERT, potentially yielding improved results.

The second category of discrepancies involves cases where the actual values are small but are erroneously predicted as large values by the models. We specifically examined three such data points in Figure~\ref{class}, which are index 22, 45, and 44. These data points involve explicit assignments of large values to variables during initialization. This might be because sometimes we assign multiple values for various purposes during initialization, such as initializing an integer, defining a matrix dimension, or setting up a for-loop. However, this poses a challenge for CodeBERT to accurately identify the intended usage of the assigned number. For instance, if we assign a very large number to an integer variable, it does not inherently increase the simulation time. Nevertheless, if CodeBERT "misunderstands" this number as the dimension of a matrix, it may erroneously predict a significantly longer simulation time. This underscores the drawback of CodeBERT when dealing with numbers as opposed to text. It further emphasizes the potential benefits of translating the C language code into a more fundamental language that can better capture the intended semantics and improve prediction accuracy.

%% file: sections/Conclusion.tex
\section{Conclusion}
In this paper, we introduce a novel dataset composed of C language programs and their simulation times (the ``SimSeconds''), specifically designed for predicting Gem5 simulation times. Through meticulous analysis of this dataset, we discerned that factors such as memory access, iteration frequency, matrix dimensionality, and data dependency can significantly influence the simulation time on Gem5. Moreover, drawing inspiration from the CodeBERT model, we constructed three distinct predictive models: a combination of CodeBERT embedding with Support Vector Regression, a CodeBERT-based Regression model, and a CodeBERT-based Classification model. Our most accurate regression model achieves a Mean Absolute Error (MAE) of 0.546, and our classification model boasts an impressive accuracy of 0.696. We hope that our proposed method can stimulate further research and lead to innovative and generic solutions.

In the future, there are a lot of possibilities to make improvements in this area. We conducted a small-scale research with a limited dataset and simulation configuration. This can be addressed in the future by expanding the dataset and conducting experiments on a wide variety of Gem5 simulation architectures. Furthermore, our present configuration is not compatible with McPAT~\cite{mcpat}. We anticipate that our methodology could also be employed to predict CPU power consumption within McPAT. Consequently, we aim to explore this possibility in future investigations, utilizing configurations that are supported by McPAT.